\documentstyle[12pt]{article}
\catcode`\@=11
\renewcommand{\thefootnote}{\fnsymbol{footnote}}

\begin{document}

\begin{titlepage}

{\hfill DFPD 95/TH/25}

{\hfill hep-th/9506102}

\vspace{1cm}

\centerline{\large{\bf INSTANTONS AND RECURSION RELATIONS}}

\vspace{0.5cm}

{\centerline{\large{\bf IN N=2 SUSY GAUGE
THEORY}\footnote[3]{Partly
supported by the European Community Research
Programme {\it Gauge Theories, applied supersymmetry and quantum
gravity}, contract SC1-CT92-0789}}}

\vspace{1.5cm}

{\centerline{\sc MARCO MATONE}}

\vspace{1cm}

\centerline{\it Department of Physics ``G. Galilei'' - Istituto Nazionale di
Fisica Nucleare}
\centerline{\it University of Padova}
\centerline{\it Via Marzolo, 8 - 35131 Padova, Italy}

\vspace{2cm}

\centerline{\sc ABSTRACT}

We find the transformation properties of the prepotential ${\cal F}$
of $N=2$ SUSY gauge theory with gauge group $SU(2)$. In particular we show that
${\cal G}(a)=\pi i\left({\cal F}(a)-{1\over 2}a\partial_a{\cal F}(a)\right)$
is modular invariant. This function satisfies
the non-linear differential equation
$\left(1-{\cal G}^2\right){\cal G}''+{1\over 4}a
{{\cal G}'}^3=0$, implying that the instanton contribution are
determined by recursion relations. Finally, we find $u=u(a)$ and
give the explicit
expression of ${\cal F}$ as function of $u$.
These results can be extended to more general cases.

\end{titlepage}

\newpage

\setcounter{footnote}{0}

\renewcommand{\thefootnote}{\arabic{footnote}}

\noindent
{\bf 1.} Recently the low-energy limit of $N=2$ super
Yang-Mills theory with gauge group $G=SU(2)$
has been solved exactly \cite{SW1}. This result has been generalized to
$G=SU(n)$ in \cite{AFKLTY} whereas the
large $n$ analysis has been investigated in \cite{DouglasShenker}.
Other interesting results concern the generalization to $SO(2n+1)$
\cite{DanSun} and non-locality at the cusp points
in moduli spaces \cite{ArgyresDouglas}.

The low-energy effective action $S_{eff}$ is derived from a single
holomorphic function ${\cal F}\left(\Phi_k\right)$ \cite{S}
\begin{equation}
S_{eff}={1\over 4\pi}{\rm Im}\left(\int d^2\theta d^2\bar\theta
\Phi^i_D\overline \Phi_i+{1\over 2}\int d^2\theta
\tau^{ij}W_iW_j\right),
\label{1}\end{equation}
where $\Phi^i_D\equiv {\partial {\cal F}/\partial \Phi_i}$ and
$\tau^{ij}\equiv {\partial^2 {\cal F}/\partial \Phi_i \partial \Phi_j }$.
Let us denote by $a_i\equiv \langle \phi^i\rangle$ and
$a^i_D\equiv \langle \phi_D^i\rangle$
the vevs of the scalar component of the chiral
superfield.
For $SU(2)$ the moduli space of quantum vacua, parametrized by
$u=\langle{\rm tr}\, \phi^2\rangle$, is
the Riemann sphere with
punctures at $u_1=-\Lambda, u_2=\Lambda$ (we will set $\Lambda=1$)
and $u_3=\infty$ and a ${\bf Z}_2$ symmetry
acting by $u\leftrightarrow -u$.
The asymptotic expansion of the prepotential has the structure
\cite{SW1}
\begin{equation}
{\cal F}={i\over 2\pi}a^2\log a^2+\sum_{k=0}^\infty {\cal F}_k
a^{2-4k}.
\label{hfgty}\end{equation}

In \cite{SW1}
the vector
$\left(a_D,a\right)$
has been considered as
a holomorphic section of a flat
bundle. In particular in \cite{SW1} the monodromy properties
of $\left(a_D(u),a(u)\right)$ have been identified with $\Gamma(2)$
\begin{equation}
\left(\begin{array}{c} a_D\\ a
\end{array}\right)\Longrightarrow
\left(\begin{array}{c} \tilde
a_D\\ \tilde a \end{array}\right)=
M_{u_i}\left(\begin{array}{c} a_D\\ a \end{array}\right),
\qquad i=1,2,3,
\label{2}\end{equation}
where
$$
M_{-1}=
\left(\begin{array}{c}-1\\-2
\end{array}\begin{array}{cc}2\\3\end{array}\right),
\quad M_{1}=
\left(\begin{array}{c}1\\ -2
\end{array}\begin{array}{cc}0\\1\end{array}\right),\quad
M_{\infty}=
\left(\begin{array}{c}-1\\0
\end{array}\begin{array}{cc}2\\-1\end{array}\right).
$$
The asymptotic behaviour of this section, derived in \cite{SW1},
 and the geometrical
data above completely determine
$\left(a_D(u),a(u)\right)$.
In particular the explict expression of the section $(a_D,a)$
has been obtained  by first constructing tori
parametrized by $u$ and then identifying a suitable meromorphic
differential \cite{SW1}.

Before considering the framework of uniformization theory,
we find the explicit expression of
${\cal F}$ in terms of $u$. Next we will find the modular properties of
${\cal F}$ by solving a linear differential equation which arises from
defining properties. We will use uniformization theory in order to
explicitly find $u=u(a)$ and to derive the (non-linear) differential
equation satisfied by ${\cal F}$ as a function of $a$. This equation
furnishes, as expected,  recursion relations which determine the
instanton contributions to ${\cal F}$. Our general formula
is in agreement with the results in \cite{KLT} where
the first six terms
of the instanton contribution have been computed.

Let us start with the explicit expression of ${\cal F}$ as function of
$u$. Let us recall that \cite{SW1}
\begin{equation}
a_D={\sqrt 2\over \pi}\int_1^u {dx \sqrt{x-u}\over \sqrt{x^2-1}},
\qquad a={\sqrt 2\over \pi}\int_{-1}^1 {dx \sqrt{x-u}\over
\sqrt{x^2-1}}.
\label{14}\end{equation}
In order to solve the problem we use the integrability of the
$1$-differential
\begin{equation}
\eta(u)=a\partial_u a_D-a_D\partial_u a={1\over \pi^2}\int_1^udx\int_{-1}^1dy
{y-x\over \sqrt{(x^2-1)(x-u)(y^2-1)(y-u)}}.
\label{51}\end{equation}
We have
\begin{equation}
g(u)=\int_1^udz\eta(z)=
{1\over \pi^2}
\int^u_1dx\int_{-1}^1dy{y-x\over \sqrt{(x^2-1)(y^2-1)}}
\log\left[{2u-x-y+2\sqrt{(u-x)(u-y)}\over x-y}\right].
\label{52}\end{equation}
On the other hand notice that
$$
\partial_u{\cal F}=a_D\partial_ua=
{1\over 2}[\partial_u(aa_D)-\eta(u)],
$$
so that, up to an additive constant, we have
\begin{equation}
{\cal F}(a(u))=
{1\over 2\pi^2}
\int^u_1dx\int_{-1}^1dy{4\sqrt{(x-u)(y-u)}-(y-x)
\log\left[{2u-x-y+2\sqrt{(u-x)(u-y)}\over x-y}\right]
\over \sqrt{(x^2-1)(y^2-1)}}.
\label{53}\end{equation}
Later, in the framework of uniformization theory, we will show that
$\eta$ is a constant (in the $u$-patch), so that $g$ is proportional to $u$.

We now find the transformation properties
of ${\cal F}(a)$. By (\ref{3}), we have
\begin{equation}
{\partial^2 \widetilde{\cal F}(\tilde a)\over \partial \tilde a^2}=
{A{\partial^2{\cal F}(a)\over \partial a^2}+B\over
C{\partial^2{\cal F}(a)\over \partial a^2}+D},
\label{54}\end{equation}
where
$\left(\begin{array}{c}A\\C
\end{array}\begin{array}{cc}B\\D\end{array}\right)
\in\Gamma(2)$ and $\tilde a=Ca_D+Da$. On the other hand
\begin{equation}
{\partial^2 \widetilde{\cal F}(\tilde a)\over \partial \tilde a^2}=
\left[ -\left({\partial \tilde a\over \partial a}\right)^{-3}
{\partial^2 \tilde a\over \partial a^2}{\partial \over \partial a}
+\left({\partial \tilde a\over \partial a}\right)^{-2}
{\partial^2\over \partial a^2}\right]
\widetilde{\cal F}(\tilde a).
\label{55}\end{equation}
Eqs.(\ref{54}) (\ref{55}) imply that
\begin{equation}
(C{\cal F}^{(2)}+D)\partial_a^2 \widetilde{\cal F}(\tilde a)-
C{\cal F}^{(3)}\partial_a \widetilde{\cal F}(\tilde a)-(A{\cal F}^{(2)}+B)
(C{\cal F}^{(2)}+D)^2=0,
\label{56}\end{equation}
where ${\cal F}^{(k)}\equiv \partial_a^k{\cal F}(a)$, whose
solution is
\begin{equation}
\widetilde
{\cal F}(\tilde a)={\cal F}(a)+{AC\over 2} a_D^2+{BD\over 2} a^2+BC aa_D.
\label{57}\end{equation}
This means that the function
\begin{equation}
{\cal G}(a)=\pi i
\left({\cal F}(a)-{1\over 2}a\partial_a{\cal F}(a)\right)=
-{\pi i\over 2}g(u),
\label{58}\end{equation}
is modular invariant, that is
\begin{equation}
\widetilde {\cal G} (\tilde a)={\cal G}(a).
\label{59}\end{equation}
By (\ref{hfgty}) we have asymptotically
\begin{equation}
{\cal G}=\sum_{k=0}^\infty {\cal G}_k a^{2-4k},
\qquad {\cal G}_0={1\over 2},\quad
{\cal G}_k=2 \pi i k{\cal F}_k.
\label{abas}\end{equation}

\vspace{0.5cm}

\noindent
{\bf 2.} In order to find $u=u(a)$ and ${\cal F}$ as function
of $a$, we need few facts about uniformization theory.
Let us denote by
$\widehat{\bf C}\equiv {\bf C}\cup\left\{\infty\right\}$ the Riemann sphere
and by $H$ the upper half plane endowed with the Poincar\'e
metric $ds^2=|dz|^2/({\rm Im}\,z)^2$.
It is well known that  $n$-punctured spheres
$\Sigma_n\equiv \widehat {\bf C}\backslash\{u_1,\ldots,u_n\}$, $n\ge 3$,
can be represented as
$H/\Gamma$ with $\Gamma\subset PSL(2,{\bf R})$ a parabolic (i.e.
with $|{\rm tr}\,\gamma|=2$, $\gamma\in\Gamma$) Fuchsian group. The map
$J_H: H\to \Sigma_n$ has the property $J_H(\gamma\cdot z)=
J_H(z)$, where $\gamma\cdot z=(Az+B)/(Cz+D)$,
$\gamma=\left(\begin{array}{c}A\\C
\end{array}\begin{array}{cc}B\\D\end{array}\right)
\in\Gamma$. It
follows that after winding around nontrivial loops the inverse map
transforms as
\begin{equation}
J_H^{-1}(u)\longrightarrow
\widetilde J_H^{-1}(u)=
{{A J_H^{-1}(u)+B}\over CJ_H^{-1}(u) +D}.
\label{3}\end{equation}
The projection of the Poincar\'e metric onto
$\Sigma_n\cong H/\Gamma$ is
\begin{equation}
ds^2=e^\varphi |du|^2=
{|{J_H^{-1}(u)}'|^2\over({\rm Im}\,
J_H^{-1}(u))^2}|du|^2,\label{5}\end{equation}
which is invariant under $SL(2,{\bf R})$ fractional transformations
of $J_H^{-1}$. The fact that $e^\varphi$ has constant curvature $-1$
means that $\varphi$ satisfies the Liouville equation
\begin{equation}
\partial_u\partial_{\bar u}\varphi={e^\varphi\over 2}.
\label{6}\end{equation}
Near a puncture we have $\varphi\sim
-\log\left(|u-u_i|^2\log^2|u-u_i|\right)$.
For the Liouville
stress tensor we have the following equivalent expressions
\begin{equation}
T(u)=
\partial_u\partial_u\varphi-{1\over 2}\left(\partial_u\varphi\right)^2=
\left\{J_H^{-1},u\right\}=\sum_{i=1}^{n-1}\left({1\over 2(u-u_i)^2}+
{c_i\over u-u_i}\right).
\label{7}\end{equation}
where $\left\{J_H^{-1},u\right\}$ denotes the Schwarzian derivative
of $J_H^{-1}$ and
the $c_i$'s, called accessory parameters, satisfy the constraints
\begin{equation}
\sum_{i=1}^{n-1}c_i=0,\qquad \sum_{i=1}^{n-1}c_iu_i=1-{n\over 2}.
\label{8b}\end{equation}

Let us now consider the covariant operators
 introduced in the
formulation of the KdV equation in higher genus \cite{BM}.
We use ${1/{J_H^{-1}}'}$ as covariantizing polymorphic vector
field \cite{M}
\begin{equation}
{\cal S}^{(2k+1)}_{J_H^{-1}}=(2k+1)
{{J_H^{-1}}'}^k \partial_u
{1\over {{J_H^{-1}}'}}
\partial_u
{1\over {{J_H^{-1}}'}}
\ldots
\partial_u
{1\over {{J_H^{-1}}'}}
\partial_u
{{J_H^{-1}}'}^k,
\label{cvprtr}\end{equation}
where the number of derivatives is $2k+1$ and $'\equiv\partial_u$.
Univalence of ${J_H^{-1}}$ implies holomorphicity
of ${\cal S}^{(2k+1)}_{J_H^{-1}}$.
An interesting property of the equation
${\cal S}^{(2k+1)}_{J_H^{-1}}\cdot \psi=0$
is that
its projection on $H$ reduces to the trivial equation
$(2k+1){z'}^{k+1}\partial_z^{2k+1}\widetilde\psi=0$,
where $z=J_H^{-1}(u)$.
Operators
${\cal S}^{(2k+1)}_{J_H^{-1}}$
are covariant, holomorphic
and $SL(2,{\bf C})$ invariant, which by (\ref{3}) implies
singlevaluedness of ${\cal S}^{(2k+1)}_{J_H^{-1}}$.
Furthermore, M\"obius invariance of the
Schwarzian derivative implies
that
${\cal S}^{(2k+1)}_{J_H^{-1}}$ depends
on $J_H^{-1}$ only through the stress tensor (\ref{7})
and its derivatives.
For $k=1/2$, we have the {\it uniformizing equation}
\begin{equation}
\left({J_H^{-1}}'\right)^{1\over 2}\partial_u{1\over {J_H^{-1}}'}
\partial_u \left({J_H^{-1}}'\right)^{1\over 2}\cdot\psi=
\left(\partial^2+{T\over 2}\right)\cdot \psi=0,
\label{9}\end{equation}
that, by construction, has the two linearly independent solutions
\begin{equation}
\psi_1=
\left({J_H^{-1}}'\right)^{-{1\over 2}}J_H^{-1},\qquad
\psi_2=
\left({J_H^{-1}}'\right)^{-{1\over 2}},
\label{minusonehalfdifferentials}\end{equation}
so that
\begin{equation}
J_H^{-1}=\psi_1/\psi_2.
\label{10}\end{equation}
By (\ref{3}) and (\ref{minusonehalfdifferentials}) it follows that
\begin{equation}
\left(\begin{array}{c} \psi_1 \\ \psi_2
\end{array}\right)\longrightarrow
\left(\begin{array}{c} \widetilde
\psi_1\\ \widetilde \psi_2 \end{array}\right)=
\left(\begin{array}{c}A\\C
\end{array}\begin{array}{cc}B\\D\end{array}\right)
\left(\begin{array}{c} \psi_1 \\ \psi_2
\end{array}\right).
\label{2b21}\end{equation}

In the case of $\Sigma_3\cong H/\Gamma(2)$, Eq.(\ref{8b}) gives
$c_1=-c_2=1/4$ and the uniformizing equation (\ref{9})
becomes\footnote{This equation has been considered also
in \cite{CerDaFe}.}
\begin{equation}
\left(\partial_u^2 +{3+u^2\over 4(1-u^2)^2}\right)\psi=0,
\label{11}\end{equation}
which is solved by Legendre functions
\begin{equation}
\psi_1=\sqrt{1-u^2}P_{-1/2},\qquad
\psi_2=\sqrt{1-u^2}Q_{-1/2}.
\label{12}\end{equation}
These solutions define
a holomorphic section
that by (\ref{2b21}) has monodromy $\Gamma(2)$.

In order to find $(a,a_D)$ we observe that by (\ref{minusonehalfdifferentials})
$\psi_1$ and $\psi_2$ are (polymorphic)
$-1/2$-differentials whereas both $a_D$ and $a$
are $0$-differentials. This fact and the asymptotic behaviour of
$(a_D,a)$ given in \cite{SW1} imply that
\begin{equation}
\left(\begin{array}{c} \psi_1\\ \psi_2 \end{array}\right)=
\left(\begin{array}{c}\sqrt{1-u^2}\partial_u a_D\\
\sqrt{1-u^2}\partial_u a \end{array}\right),
\label{13}\end{equation}
where $\sqrt{1-u^2}$ is considered as a $-3/2$-differential.
Comparing with (\ref{12}) we get (\ref{14}).

\vspace{0.5cm}

\noindent
{\bf 3.} By Eqs.(\ref{11}) and (\ref{13}) it follows
that $a_D$ and $a$  are solutions of the third-order equation
\begin{equation}
\left(\partial_u^2 +{3+u^2\over
4(1-u^2)^2}\right)\sqrt{1-u^2}\partial_u\phi=0.
\label{11bis}\end{equation}
 Let us consider some aspects of this equation.
First of all note that, as observed in \cite{KLT},
\begin{equation}
\left(\partial_u^2 +{3+u^2\over
4(1-u^2)^2}\right)\sqrt{1-u^2}\partial_u\phi=
{1\over \sqrt{1-u^2}}\partial_u\left[(1-u^2)\partial_u^2-{1\over 4}
\right]\phi=0.
\label{61}\end{equation}
It follows that $\left[(1-u^2)\partial_u^2-{1\over 4}
\right]\phi=c$ with $c$ a constant. A check shows
that $a_D$ and $a$ in (\ref{14})
satisfy this equation with $c=0$
\begin{equation}
\left[(1-u^2)\partial_u^2-{1\over 4}
\right]a_D=\left[(1-u^2)\partial_u^2-{1\over 4}
\right]a=0.
\label{jdjd}\end{equation}
As noticed
in \cite{KLT},
this explains also why, despite of the fact that $a$ and $a_D$ satisfy
the third-order differential equation (\ref{11bis}),
they have two-dimensional
monodromy.
Eq.(\ref{jdjd})
is the crucial one to find $u=u(a)$ and to determine the
instanton contributions. In our framework the problem of finding the
form of ${\cal F}$ as a function of $a$ is equivalent to the following
general basic problem which is of interest also from a mathematical point of
view:

{\it Given a second-order differential equation with solutions $\psi_1$ and
$\psi_2$ find the function ${\cal F}_1(\psi_1)$
(${\cal F}_2(\psi_2)$) such that $\psi_2=\partial{\cal F}_1/\partial
\psi_2$
($\psi_1=\partial{\cal F}_2/\partial
\psi_2$).}

 We show that such a function satisfies a non-linear
differential equation.
The first step is to observe that by
(\ref{jdjd}) it follows that
\begin{equation}
aa_D'-a_Da'=c.
\label{jdsuw}\end{equation}
Since $(a_D,a)$ are
(polymorphic) $0$-differentials, it follows that in changing patch
the constant $c$
in (\ref{jdsuw})
is multiplied by the Jacobian of the coordinate transformation.
Another equivalent way to see this, is to notice
that Eq.(\ref{jdjd})
gets a first derivative
under a coordinate transformation.
Therefore in another patch the r.h.s. of (\ref{jdsuw}) is no longer a
constant.
As we have seen, covariance of the equation such has
$$
(\partial^2_z+F(z)/2)\psi(z)=0,
$$ is ensured if and only if
$\psi$ transforms as a $-1/2$-differential and $F$ as a Schwarzian
derivative. In terms of the solutions
$\psi_1$, $\psi_2$ one can construct the $0$-differential
$\psi_1'\psi_2-\psi_1\psi_2'$ that, by the structure of the equation,
is just a constant $c$. In another patch we have
$(\partial^2_w+\widetilde F(w)/2)\widetilde\psi(w)=0$,
so that $\psi_1(z)\partial_z\psi_2(z)-\psi_2(z)\partial_z\psi_1(z)=
\widetilde\psi_1(w)
\partial_w\widetilde\psi_2(w)-
\widetilde\psi_2(w)\partial_w\widetilde\psi_1(w)=c$.

This discussion shows that flatness of
$a_D$ and $a$ is the reason
of the reduction mechanism from the third-order to second-order
equation.

By (\ref{51}) (\ref{52}) (\ref{58})  and (\ref{jdsuw}) it follows that
\begin{equation}
Au+B={\cal G}(a),
\label{0ofcd}\end{equation}
where $B$ is a constant which we will show to be zero. To determine the
constant $A$, we note that asymptotically $a\sim \sqrt {2u}$, therefore
by (\ref{abas}) one has $A=1$.
By (\ref{14}) and (\ref{0ofcd})
it follows that
\begin{equation}
a_D={\sqrt 2\over \pi}\int_1^{{\cal G}(a)+B}
{dx \sqrt{x-{\cal G}(a)-B}\over \sqrt{x^2-1}},
\qquad a={\sqrt 2\over \pi}\int_{-1}^1 {dx \sqrt{x-{\cal G}(a)-B}\over
\sqrt{x^2-1}}.
\label{14bis}\end{equation}
Apparently to solve these two equivalent
integro-differential equations seems a difficult task. However we
can use the following trick. First notice that
\begin{equation}
\left[(1-u^2)\partial_u^2-{1\over 4}\right]\phi=0=
\left\{\left[1-({\cal G}+B)^2\right]
\left({\cal G}'\partial_a^2-{\cal G}''\partial_a\right)-
{1\over 4} {{\cal G}'}^3\right\}\phi=0,
\label{hfgty1}\end{equation}
where now $'\equiv \partial_a$.
Then, since $\phi=a$ (or equivalently $\phi=a_D=\partial_a{\cal F}$)
is a solution of (\ref{hfgty1}),
it follows that ${\cal G}(a)$ satisfies the non-linear differential
equation
$\left[1-({\cal G}+B)^2\right]{\cal G}''+{1\over 4}a
{{\cal G}'}^3=0$.
Inserting the expansion
 (\ref{abas}) one can check that the only way to compensate the
$a^{-2(2k+1)}$ terms is to set $B=0$. Therefore
\begin{equation}
\left(1-{\cal G}^2\right){\cal G}''+{1\over 4}a
{{\cal G}'}^3=0,
\label{gdfet}\end{equation}
which is equivalent to the following recursion relations
for the instanton contribution
(recall that
${\cal G}=2\pi i k {\cal F}_k$)
$$
{\cal G}_{n+1}={1\over 8{\cal G}_0^2(n+1)^2}\cdot
$$
\begin{equation}
\cdot\left\{
(2n-1)(4n-1){\cal G}_n
+2{\cal G}_0
\sum_{k=0}^{n-1}{\cal G}_{n-k}{\cal G}_{k+1}c(k,n)
-2\sum_{j=0}^{n-1}\sum_{k=0}^{j+1}{\cal G}_{n-j}
{\cal G}_{j+1-k}{\cal G}_{k}d(j,k,n)\right\},
\label{recursion2}\end{equation}
where $n\geq 0$, ${\cal G}_0=1/2$ and
$$
c(k,n)=2k(n-k-1)+n-1,
\qquad
d(j,k,n)=
[2(n-j)-1][2n-3j-1+2k(j-k+1)].
$$
The first few terms are
${\cal G}_0={1\over 2},\; {\cal G}_1={1\over 2^2},\;
{\cal G}_2={5\over 2^6},\; {\cal G}_3={9\over 2^7},$
in agreement\footnote{Notice that we are using
different normalizations, thus to compare with
${\cal F}_k^{KLT}$ in \cite{KLT}
one should check the invariance of the quantity
${{\cal F}_k\over {\cal F}_k^{KLT}}{{\cal F}_{k+1}^{KLT}
\over {\cal F}_{k+1}}$.} with the results in \cite{KLT} where
the first terms of the instanton contribution
have been computed by first inverting $a(u)$ as a series for large
$a/\Lambda$ and then inserting this in $a_D$.

Finally let us notice that the inverse of $a=a(u)$ is
\begin{equation}
u={\cal G}(a),
\label{01}\end{equation}
and
\begin{equation}
aa_D'-a_Da'={2i\over \pi},
\label{02}\end{equation}
which is useful to explicitly determine the critical curve on which
${\rm Im}\, a_D/a=0$, whose
structure has been considered in \cite{SW1}\cite{F}\cite{AFS}.

\vspace{0.5cm}

\noindent
{\bf Acknowledgements.} We would like to thank
P. Argyres, F. Baldassarri,
G. Bonelli, J. de Boer, J. Fuchs, W. Lerche,
P.A. Marchetti, P. Pasti and M. Tonin for useful discussions.

\end{document}